# HUBBLE AND SHAPLEY - TWO EARLY GIANTS OF OBSERVATIONAL COSMOLOGY

National Research Council of Canada
Dominion Astrophysical Observatory
Herzberg Institute of Astrophysics
5071 West Saanich Road, Victoria, British Columbia, V9E 2E7, Canada
Sidney van den Bergh  Sidney.vandenbergh@nrc.ca)

ABSTRACT

Observational cosmology of the first decades of the Twentieth Century was dominated by two giants: Edwin Hubble and Harlow Shapley. Hubble's major contributions were to the study and classification of individual galaxies with large telescopes, whereas Shapley is best remembered for his work on groups and clusters of galaxies using telescopes of more modest aperture.

Harlow Shapley was a gregarious man who liked telling stories, was a fine dinner companion, and obviously enjoyed having lived the American dream. Born on a hay farm in the Ozarks he ended up as the director of the Harvard College Observatory. I never met Hubble, but from his biography (Christianson 1995) he sounds like a rather remote character with a psuedo Oxford accent. Hubble graduated from Oxford in 1913, just as the British Empire reached its apogee, and before it started to bleed to death during the Great War. In his autobiography Shapley (1969, p. 57) writes, "Hubble just didn't like people. He didn't associate with them, didn't care to work with them". Hubble had a sharp legal mind which he applied with great success to the problems of observational cosmology. A friend of Hubble's once wrote, "After scoring a debating point, he would light his pipe, flip the match into the air so that it described a circle, and catch it, still burning as it came down. "Like a good lawyer he was able to use observations to present particularly convincing arguments. Good examples are Hubble (1936, p. 118) where he simultaneously plots the images of galaxies and their spectra to illustrate the velocity-distance relationship, and the "tuning fork" diagram (Hubble 1936, p. 45) which beautifully illustrates the transition from elliptical through S0 to spiral morphologies. He did this in such a convincing way that it was not noticed for half a century that S0 galaxies are typically only half as luminous as E and Sa galaxies. Thus providing strong evidence against his speculative hypothesis that lenticular galaxies constitute an intermediate evolutionary stage between elliptical and spiral galaxies. Although we now associate Hubble's name with the tuning fork diagram Block & Freeman (2008) have shown quite convincingly that these ideas actually originated with J. H. Reynolds and others.

Edwin Hubble (1889-1953) and Harlow Shapley (1885-1972) were both born in Missouri and recognized as rising stars by George Ellery Hale, who hired them to work at the Mt. Wilson Observatory where Hubble remained for his entire career, while Shapley left for Harvard in

1921. According to Christianson (1995, p.149) " Hubble was privately anxious for Shapley to go. He was not accustomed to standing in anyone else's shadow, especially one whose conduct he considered obnoxious and whose ambitions matched his own." For his outstanding work Shapley was awarded the RAS Gold medal in 1934, Hubble received this same medal in 1940. Later Hubble was awarded the Bruce Gold medal of the ASP in 1938. Shapley received that award in 1939. Both men will primarily go down in history for a single great discovery. Shapley (1918) for using the distribution of Galactic globular clusters to show that the center of our Milky Way system is located far from the Sun in the direction of Sagittarius. The importance of this transition from a heliocentric to a Galactocentric paradigm was comparable to the change from a geocentric to a heliocentric Universe proposed by Copernicus in 1543. Shapley (1969, p. 60) himself regarded this discovery that the Sun (and hence mankind) was peripheral, rather than central, as one of the most important thoughts that he had ever had. Hubble's name will forever be associated with the discovery in 1923 of Cepheid variables in the great Andromeda nebula, which resolved the longstanding debate about the nature of spiral galaxies. Here there is also a link to Shapley (1914) who had previously hatched the hypothesis that Cepheids were pulsating variables, rather than binaries, as had previously been thought.

The careers of these two scientists diverged with Hubble's main contributions involving the classification and distance determination of individual galaxies using the world's largest telescopes. On the other hand Shapley is perhaps best remembered for his work on the distribution and clustering of galaxies, which he studied with wide-field telescopes of relatively modest aperture. Hubble's longest (but perhaps least influential paper) concerned the large-scale distribution of galaxies over the sky. From deep studies of over one thousand small fields thinly distributed over the sky north of declination -30$^o$ Hubble (1934) found that (1) absorption by dust causes a deficit of galaxies at low Galactic latitudes and (2) the distribution of galaxies on very large scales is homogeneous, i.e. there is no indication of a super system of nebulae. Perhaps his deepest insight (Hubble 1936, p. 82) into the nature of galaxy clustering is contained in the statement that "The groups (such as the Local Group) are aggregations drawn from the general field, and are not additional colonies superposed on the field." Hubble (1936), p. 77 also adds that "Pending definite information, it is supposed that the frequency diminishes as the population increases, over the whole range of groups and loose clusters to the great clusters themselves. "Because Hubble's survey sampled only widely distributed very small fields, it was not suitable for the study of the large-scale distribution of distant clusters of galaxies. Nor could Hubble's sparsely sampled data provide much information on the structure and distribution of chains and clusters of galaxies across the heavens. On the other hand Shapley's wide-field and all-sky surveys were particularly well-suited to the study of galaxy clusters and their distribution over the sky. Perhaps Shapley's his greatest contribution to astronomy was the Shapley-Ames (Shapley, & Ames 1932a,b) catalog of the brightest galaxies in the sky, which was mainly based on observations with small wide-field telescopes. On the other hand Hubble is probably best remembered for his classification system for galaxies and for determining the distances to the nearest galaxies using the Mt Wilson 100-in. telescope. Plots of the projected distribution of galaxies in the Shapley-Ames catalog over the sky provided three profound new insights: (1) There is no concentration of galaxies towards our own Milky Way system, (2) galaxies are distributed in a very clumpy fashion with major flattened clusterings - such as the Virgo super cluster, and (3) Shapley & Ames (1932a) discovered what they referred to as "high latitude vacancies" -structures that are nowadays referred to as voids. Regarding the discovery of such voids Shapley & Ames (1932b) write: "The vacancies [in the distribution of galaxies] are important. They are not due to the insufficiency of the survey, for over the whole sky the search has been thorough and the plates adequate. Nor are the barren regions, such as that at $\lambda = 20^o$, $b = + 50^o$, the result of heavy obscuration, as in low latitudes, since nebulae fainter than the thirteenth magnitude appear in average abundance in these high latitude regions. "Shapley (1930) was also able to show that clusters of galaxies are not distributed at random, but are concentrated in super clusters. The greatest of these super clusters in nearby regions of the Universe is the so called Shapley (1930) concentration [see Proust et al. (2006)]. An intrinsic limitation on Shapley's studies of the large-scale structure of the Universe was that his data only provided two- dimensional mapping of the distribution of clusters over the sky. The first tentative steps towards study of their three-dimensional distribution were based on the studies of the radial velocities of large numbers of galaxies at intermediate distances by Gregory & Thomson (1978) and Joeveer, et al. (1978). Our present understanding of the nature of the large-scale distribution of galaxies in the Universe resulted from the large and homogeneous three-dimensional mapping by de Lapparent et al. (1986). This work revealed that galaxies are distributed throughout the Universe in enormous frothy bubbles and chain-like structures. The properties of dark matter particles determine the strucutre of this cosmic web.

Finally it is of interest to note that Hubble's galaxy classification system excluded both the most luminous galaxies (quasars) and the most numerous objects – dim spheroidal galaxies. The first of these dwarf spheroidals, the Sculptor and Fornax systems, were discovered by Shapley (1938).

In Hubble's obituary Humason (1954) wrote: He was sure of himself -of what to do, and how to do it. " On the other hand Shapley's (1969) motto seems to have been "I always worry". They were clearly very different men.

I am indebted to David Block, Bonnie Bullock, Brenda Parrish and Virginia Trimble for their kind help.

References

Block, D. L. & Freeman, K. C. (2008). Shrouds of the Night (pp.199-213) Berlin: Springer.

Christianson, G. E. (1995). Mariner of the Nebulae .Toronto: Harper Collins Canada Ltd

de Lapparent, V., Geller, M. J. & Huchra, J. P. (1986). A Slice of the Universe. ApJ. 302, L1

Gregory, S. A. & Thompson, L. A. (1978). The Coma/A1367 Supercluster .and its Environs, ApJ. 222, 784

Hubble, E. (1934). The Distribution of Extra-galactic Nebulae, ApJ, 79, 8

Hubble, E. (1936). The Realm of the Nebulae. New Haven: Yale University Press.

Humason, M. L. (1954). Edwin Hubble, MNRAS, 114, 29

Joeveer, M., Einasto, M & Tago,E. (1978). Spatial Distribution of Galaxies and clusters in the Southern Hemisphere. MNRAS, 185, 357

Proust, D. et al. (2006). The Shapley Supercluster. The Largest Concentration in the Local Universe. ESO Messenger, No. 124, p. 30

Shapley, H. (1914). On the Nature and Cause of Cepheid Variation, ApJ, 40, 448, 1914

Shapley, H. (1930). Note on a Remote Cloud of Galaxies in Centaurus Harvard Bull. 874 (p.9)

Shapley, H. (1933). On the Distribution of Galaxies. Pub. Nat. Acad, Sci. 19, 389

Shapley (1938). Two Stellar Systems of a New Kind. Nature 142, 715

Shapley (1969). Through Rugged Ways to the Stars (p.6), New York, Charles Scribner's Sons

Shapley, H. & Ames, A. (1932a), A Survey of the External Galaxies, Brighter than the Thirteenth Magnitude, Ann Harvard College Obs. 88, No. 2

Shapley, H. & Ames, A. (1932b). Photometric Survey of the Nearer Extragalactic Nebulae. Harvard Bull. 887, p.1